\documentclass[aps,prl,final,twocolumn,letterpaper]{revtex4}

\usepackage{graphicx}   
\usepackage{import}                         
\usepackage{epstopdf}
\usepackage{amsmath} 
\usepackage{float} 
\usepackage{bm}
\usepackage{amssymb}
\usepackage{quotes}
\usepackage{indentfirst}
\usepackage{color}
\usepackage{transparent}
\usepackage{dcolumn}
\usepackage{braket}
\usepackage{multirow}
\usepackage{cancel} 
\usepackage{mdframed}
\usepackage{color}
\usepackage{bm}
\usepackage{dsfont}
\usepackage{slashed}
\usepackage{soul, color} 
\soulregister\ref{7}  
\soulregister\cite{7} 
\renewcommand{\st}[1]{}

\usepackage{xr}

\makeatletter
\newcommand*{\addFileDependency}[1]{
  \typeout{(#1)}
  \@addtofilelist{#1}
  \IfFileExists{#1}{}{\typeout{No file #1.}}
}
\makeatother

\newcommand*{\myexternaldocument}[1]{%
    \externaldocument{#1}%
    \addFileDependency{#1.tex}%
    \addFileDependency{#1.aux}%
}

\myexternaldocument{si}

\usepackage{textcomp} 
\usepackage{xifthen}
\usepackage{xcolor}
\usepackage{etoolbox}
\newboolean{togglechanges} 

\setboolean{togglechanges}{false}

\newcommand{\comment}[1]{\ifbool{togglechanges}
    {#1}  
    {\textcolor{blue}{#1}}}

\usepackage{bibentry}

\usepackage{graphicx}
\usepackage{dcolumn}
\usepackage{bm}

\usepackage{textcomp} 

\begin{document}
\rmfamily

\title{Biasing the quantum vacuum to control macroscopic probability distributions}

\author{Charles~Roques-Carmes$^{\ddag,1}$}
\email{chrc@mit.edu}
\author{Yannick~Salamin$^{\ddag,1,2}$}
\email{salamin@mit.edu}
\author{Jamison~Sloan$^{1}$}
\author{Seou~Choi$^{1}$}
\author{Gustavo~Velez$^{1}$}
\author{Ethan~Koskas$^{1}$}
\author{Nicholas~Rivera$^{2,3}$}
\author{Steven~E.~Kooi$^{4}$}
\author{John~D.~Joannopoulos$^{2,4}$}
\author{Marin~Solja\v{c}i\'{c}$^{1,2}$}

\affiliation{$^\ddag$ denotes equal contribution.\looseness=-1}
\affiliation{$^{1}$ Research Laboratory of Electronics, MIT, Cambridge MA USA\looseness=-1}
\affiliation{$^{2}$ Department of Physics, MIT, Cambridge MA USA\looseness=-1}
\affiliation{$^{3}$ Department of Physics, Harvard University, Cambridge MA USA\looseness=-1}
\affiliation{$^{4}$ Institute for Soldier Nanotechnologies, MIT, Cambridge MA USA\looseness=-1}



\clearpage

\renewcommand{\sp}{\sigma_+}
\newcommand{\pbit}{$p$-bit}
\newcommand{\pbits}{$p$-bits}
\newcommand{\sm}{\sigma_-}

\vspace*{-2em}



\begin{abstract}
    One of the most important insights of quantum field theory is that electromagnetic fields must fluctuate. Even in the vacuum state, the electric and magnetic fields have a nonzero variance, leading to ubiquitous effects such as spontaneous emission, the Lamb shift, the Casimir effect, and more. These "vacuum fluctuations" have also been harnessed as a source of perfect randomness, for example to generate perfectly random photonic bits. Despite these achievements, many potential applications of quantum randomness in fields such as probabilistic computing rely on \textit{controllable} probability distributions, which have not yet been realized on photonic platforms. In this work, we show that the injection of vacuum-level “bias” fields into a multi-stable optical system enables a controllable source of “biased” quantum randomness. We demonstrate this concept in an optical parametric oscillator (OPO). Ordinarily, an OPO initiated from the ground state develops a signal field in one of two degenerate phase states (0 and $\pi$) with equal probability. By injecting bias pulses which contain less than one photon on average, we control the probabilities associated with the two output states, leading to the first controllable photonic probabilistic bit (\pbit). We shed light on the physics behind this process, showing quantitative agreement between theory and experiment. Finally, we demonstrate the potential of our approach for sensing sub-photon level fields by showing that our system is sensitive to the temporal shape of bias field pulses far below the single photon level. Our results suggest a new platform for the study of stochastic quantum dynamics in nonlinear driven-dissipative systems, and point toward possible applications in ultrafast photonic probabilistic computing, as well as the sensing of extremely weak fields.    
\end{abstract}

\maketitle

\section*{Introduction} 

One of the most fundamental and peculiar features of quantum physics is the presence of vacuum fluctuations~\cite{loudon2000quantum}: even in the apparent absence of matter or radiation, quantum fields exhibit a nonzero variance. Vacuum fluctuations of the quantum electromagnetic field underlie many intriguing phenomena, such as spontaneous emission~\cite{purcell1946}, the Lamb shift~\cite{lamb1947fine}, Casimir and van der Waals forces~\cite{kardar1999friction, chan2001quantum, sandoghdar1992direct, wilson2011observation}, as well as forms of cosmological radiation~\cite{hawking1974black, unruh1976notes}. Consequently, there has been significant interest in measuring the amplitude of electromagnetic vacuum fluctuations and their correlations, either directly, with methods such as homodyne tomography~\cite{smithey1993measurement} or electro-optic sampling~\cite{riek2015direct, benea2019electric}, or indirectly by observing their macroscopic signatures~\cite{fragner2008resolving, hoi2015probing}. 

Sensitive measurements of energy levels, forces, or fields are often required to discern the impact of vacuum fluctuations. However, certain nonlinear systems are so sensitive to initial conditions that vacuum fluctuations critically influence the outcome of some classical observable. Famously, multistable systems which spontaneously break continuous or discrete symmetries reveal the statistical nature of the vacuum~\cite{kardar1999friction}. When symmetry is broken spontaneously, the system driven by stochastic vacuum forces randomly "picks" one of the steady states:
this makes vacuum fluctuations a natural source of randomness~\cite{herrero2017quantum}. In optics, a direct and macroscopic manifestation of electromagnetic vacuum fluctuations occurs at the thresholds of lasers and optical parametric oscillators (OPOs)~\cite{haken1980synergetics, drummond1980non, drummondquantum, haken1985laser, haake1978decay}. The initial output phase of a laser is determined through fluctuation-induced spontaneous emission; on the other hand, OPOs are bistable, and vacuum fluctuations drive the system into one of two degenerate output phases with equal likelihood. As a result, these optical platforms have been thoroughly explored as generators of true and ultra-high-rate random numbers~\cite{gabriel2010generator, jofre2011true, marandi2012all,okawachi2016quantum, steinle2017unbiased, kim2021massively}. 

These quantum sources of randomness, beyond their fundamental interest, are also promising for many applications. For example, stochastic components are essential across a range of computing tasks, from probabilistic machine learning to Monte Carlo methods~\cite{ackley1985learning, gendreau2010handbook, blundell2015weight, ghahramani2015probabilistic, camsari2019p, markovic2020physics, nichol2021glide}. Probabilistic computing~\cite{camsari2019p} in particular has shown great promise for speeding up optimization and inference tasks~\cite{borders2019integer, aadit2022massively, pervaiz2018weighted, Harabi2022} while circumventing the implementation challenges of quantum computers. Its central building block is a probabilistic bit (\pbit): a stochastic logical unit described by a controllable probability distribution (e.g., producing 1 with a probability $p$, and 0 with probability $1-p$ \cite{borders2019integer}).
The prospect of \textit{leveraging vacuum fluctuations as a source of controllable randomness} for photonic probabilistic computing is very attractive, given the already-demonstrated potential advantages of photonic analog platforms in terms of computational efficiency and speed~\cite{wetzstein2020inference}.

Despite these interests, a substantial gap separates the current state-of-the-art from the proposed controlled quantum randomness in optics. Specifically, all existing optical random bit generators rely on perfect symmetry, so that vacuum fluctuations yield unbiased outcomes. The other extreme for nonlinear optical systems lies in externally seeded devices where the steady state is completely determined, sacrificing any notion of randomness. Macroscopic injection-seeding mechanisms are routinely used to enable narrow bandwidth and single-mode operation in OPOs~\cite{fix1993tunable, barnes1992injection}, microwave oscillators~\cite{kurokawa1973injection}, and lasers~\cite{stover1966locking, buczek1973laser}. Interestingly, the injection of very weak bias fields (down to sub-photon levels) is also known to influence the turn-on time of lasers~\cite{arecchi1989dynamics, littler1990detection, aquino1991nonlinear}. The existence of these two extremes --- perfect unbiased randomness and complete determinism --- raises questions about what physics may lie in between, and whether this may provide a way to harness electromagnetic vacuum fluctuations as a source of controllable ``biased'' randomness.


Here, we show that the injection of very weak ("vacuum-level") fields into a multistable optical system enables a controllable source of ``biased'' quantum randomness. Key to our work is the injection of bias fields with amplitudes comparable to that of vacuum fluctuations, enabling us to continuously interpolate between the regimes of perfect unbiased randomness and perfect determinism. We show that quantum biased randomness can be realized in various nonlinear photonic systems, and demonstrate our concept in a biased OPO, where the random variable is the phase of the signal field,  which can take on values 0 and $\pi$. We showcase the potential of our concept in three key experiments: (1) we generate the first photonic \pbit, with a parameter $p$ that can be tuned continuously from 0 to 1 through control over the bias amplitude and phase. We then elucidate the physical origin of this biased symmetry breaking by (2) showing that this biased randomness arises directly from interference between the bias field and vacuum fluctuations. Finally, probability measurements in this system enable the translation of vacuum-level perturbations into easily manipulable macroscopic observables, facilitating a new method to (3) probe vacuum field fluctuations, as well as the temporal profile of very weak electromagnetic pulses, down to much less than one ($\approx10^{-3}$) photon per pulse. Our results pave the way towards configurable macroscopic probability distributions relying on biased quantum vacuum fluctuations as random seed, and offer a playground to study intricate dynamics initiated by vacuum-level fields in nonlinear driven-dissipative quantum systems.

\begin{figure}
    \centering
    \includegraphics[scale = 0.65]{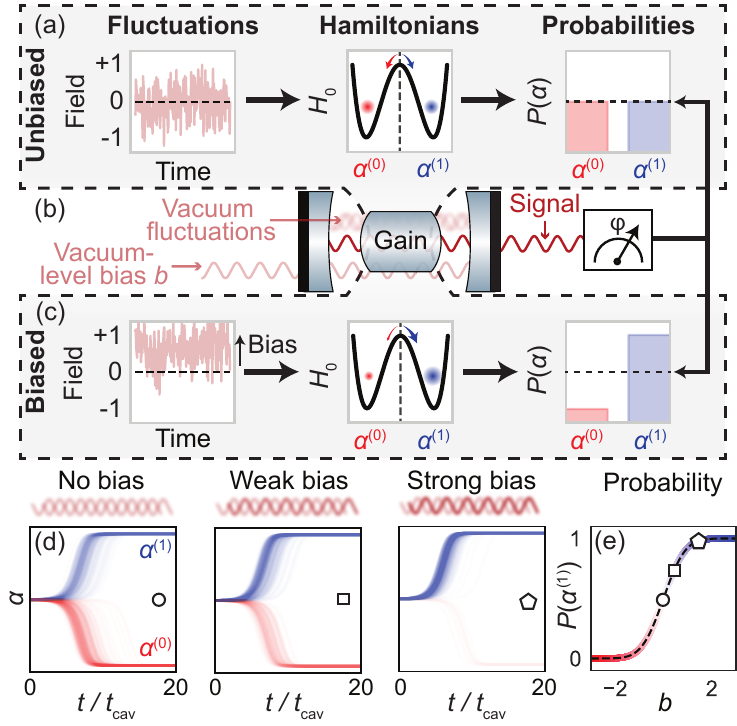}
    \caption{\textbf{Tuning the probability distribution of a multistable system by biasing vacuum fluctuations.} (a, c) Fluctuation snapshot, Hamiltonian, and associated probabilities for an (a) unbiased and (c) biased system. We consider here a bistable system for illustration, exhibiting degenerate local minima. In (c), a small bias field, whose intensity is commensurate with the vacuum fluctuations, is also injected in the cavity to control the signal's probability distribution. (b) Schematic of quantum optical multistable system. Gain is generated via nonlinear interaction in the cavity (e.g., via stimulated emission, $\chi^{(2)}$, or $\chi^{(3)}$ interaction). (d) Corresponding amplitude trajectories for a biased optical parametric oscillator (OPO). One hundred trajectories are plotted in each panel and time is normalized to the cavity lifetime $t_\text{cav}$. (e) Probability of observing state $\alpha^{(1)}$ (corresponding to OPO signal phase 0) as a function of bias intensity $b$, such that $b^2$ is the number of bias photons in the signal mode.}
    \label{fig:concept}
\end{figure}

\section*{Controlling macroscopic observables by coherent injection of vacuum-level fields in quantum optical multistable systems}

We first describe a general method to generate probability distributions encoded in macroscopic observables of nonlinear driven-dissipative physical systems, controlled by weakly biased vacuum fluctuations. We consider a physical system with Hamiltonian $H_0$, which possesses multiple degenerate stable steady states (multistable). For concreteness, consider a bistable system described by two degenerate ground states $\alpha^{(i)}$ ($i\in\{0,1\}$), as shown in Figure~\ref{fig:concept}(a). The system is initiated in the unstable state which lies at the local maximum of the potential between the two degenerate states. This unstable initial state will then decay into one of the degenerate energy minima with equal probability (Figure~\ref{fig:concept}(a), center). The resulting random outcome stems from random field fluctuations (depicted in Figure~\ref{fig:concept}(a), left) in the initial state, and is a manifestation of discrete symmetry breaking. By making repeated outcome measurements in statistically independent systems, one can construct a probability distribution associated with the multiple outcomes (Figure~\ref{fig:concept}(a), right). In a perfectly symmetric system described by $H_0$, the statistics of the ground state are trivial $P(\alpha^{(0)})=P(\alpha^{(1)})=\frac{1}{2}$. Nonlinear quantum optical multi-stable systems provide a natural platform to implement such a Hamiltonian, where macroscopic probability distributions can be extracted by repeated phase measurements (Figure~\ref{fig:concept}(b)). 

By adding a weak bias to the system, the probability distribution of outcomes can be manipulated. Although adding a coherent field is known to break the symmetry of such systems, we specifically consider vacuum-level bias fields, which are \emph{on the order of the fields associated with vacuum fluctuations}. For such vacuum-level bias fields acting on a macroscopic system, the local minima of the energy landscape are close, but not exactly degenerate. However, these vacuum-level bias fields can dramatically influence the system as it evolves from its initial condition (an unstable state at $\alpha=0$ in our example). They add up coherently to the vacuum field, slightly displacing its mean value towards one of the stable states (Figure~\ref{fig:concept}(c), left). Due to the (externally) broken symmetry, the system will evolve into one of the stable states with unequal probabilities. In the bistable case, this results in a binomial distribution $\mathcal{B}_p$ : $P(\alpha^{(0)})=1-p$ and $P(\alpha^{(1)})=p$ with $p\in[0,1]$. A generalization of this concept to multistable systems and considerations on the implementation of \pbits~in ultrafast optics are presented in the Supplementary Materials (SM), Section~S1.


In the following, we focus on one possible implementation of Figure~\ref{fig:concept}(b): a bistable optical system based on a degenerate biased OPO, which we experimentally demonstrate in this work. The essential component of a degenerate OPO is a second-order nonlinear crystal in an optical cavity. When a pump with frequency $2\omega$ is coupled into the crystal, oscillations at the signal frequency $\omega$ are induced through parametric amplification. The resulting steady state signal field can take on two distinct phases, 0 and $\pi$, respectively $\alpha^{(1)}$ and $\alpha^{(0)}$. In the absence of any injected bias field at the signal frequency, the OPO steady state above threshold approaches a mixed state of the two possible phases, with equal likelihood. The outcome is determined spontaneously as vacuum fluctuations break the discrete symmetry between the two possible states~\cite{marandi2012all}.

The story changes when a bias field is injected into the cavity. By tuning the amplitude and phase of the injected bias field, one creates a mixed state density matrix of the form $\rho=(1-p)\ket{\alpha^{(0)}}\bra{\alpha^{(0)}}+p\ket{-\alpha^{(0)}}\bra{-\alpha^{(0)}}$, realizing the binomial distribution $\mathcal{B}_p$ discussed above (with $\ket{-\alpha^{(0)}} = \ket{\alpha^{(1}}$). Since $p$ can be continuously tuned from 0 to 1, a phase measurement of this state serves as controllable $p$-bit. The probability $p$ of observing the $0$ phase ($\alpha^{(1)}$ state) is given by the analytical expression:
\begin{equation}
    p(b) = \frac{1}{2}\left[\text{erf}\left(\frac{|b|\cos\phi}{b_0}\right) + 1\right].
    \label{eq:prob-bias}
\end{equation}
In this expression, $\text{erf}$ is the error function, $b$ is the bias field which is defined such that $b^2$ is the number of bias photons present in the signal pulse in the cavity during each round trip, and $\phi$ the phase of the bias field relative to the phase of the OPO signal. Furthermore, $b_0$ is the critical bias field level required to modify the probability, which is of order unity, corresponding to single-photon level bias. Amplitude trajectories for an OPO with weak gain and various levels of bias power are shown in Figure~\ref{fig:concept}(c,d). The derivation of the bias-probability relationship from a rigorous quantum mechanical model is given in Section~S2 of the SM. Notably, these principles can be applied beyond OPOs to realize different controllable probability distributions. In Section~S1 of the SM, we show a generalization of this concept to a biased multistable system consisting of a laser with vacuum-level injection seeding, realizing a \pbit~over a continuous domain.

\begin{figure*}
    \centering
    \includegraphics[scale = 0.42]{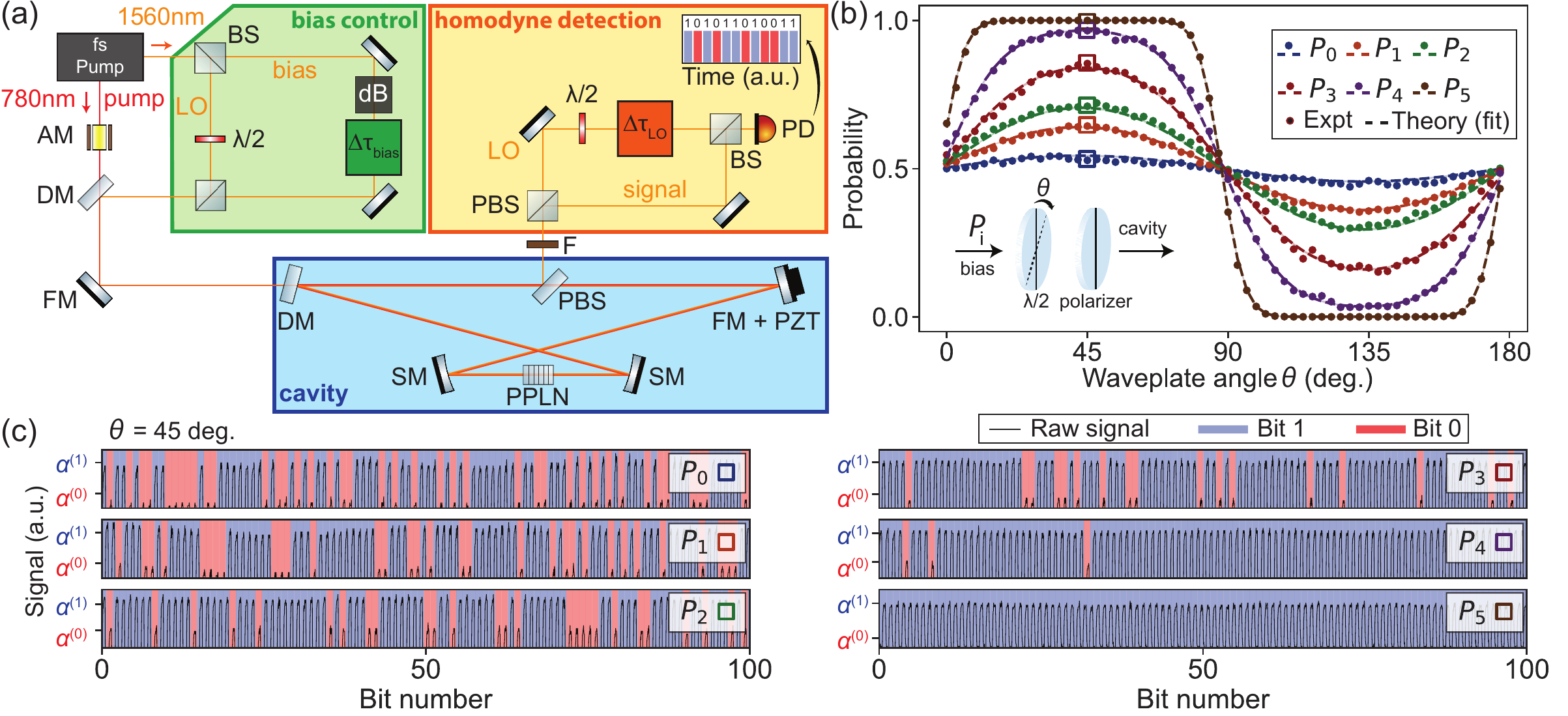}
    \caption{\textbf{Experimental demonstration of a photonic \pbit~in a biased OPO.} (a) Schematic of the experimental setup to generate and measure tunable probability distributions in an OPO. AM: amplitude modulator, (P)BS : (polarization) beam splitter, dB : bias attenuation (combination of neutral density filters, pinhole and polarization optics), $\lambda/2$ : half waveplate, DM : dichroic mirror, PD : photodiode, FM : flat mirror, SM : spherical mirror, PPLN : periodically-poled lithium niobate nonlinear crystal, LO : local oscillator, F : spectral filter, PZT: piezo-electric actuator. (b) Probability measurement for synchronous pump and bias ($\Delta\tau_\text{bias}=0$) for various maximum bias powers $P_i$ and varying waveplate angle $\theta$, over 20,000 samples. Inset: schematic of the bias control setup in this experiment, such that the bias field in the crystal is $\propto \sqrt{P_i} \sin \left(2\theta\right)$. (c) Bit stream portions at $\theta = 45^\circ$ for various incident maximum powers $P_i$.}
    \label{fig:expt}
\end{figure*}

\begin{figure}
    \centering
    \includegraphics[scale = 0.45]{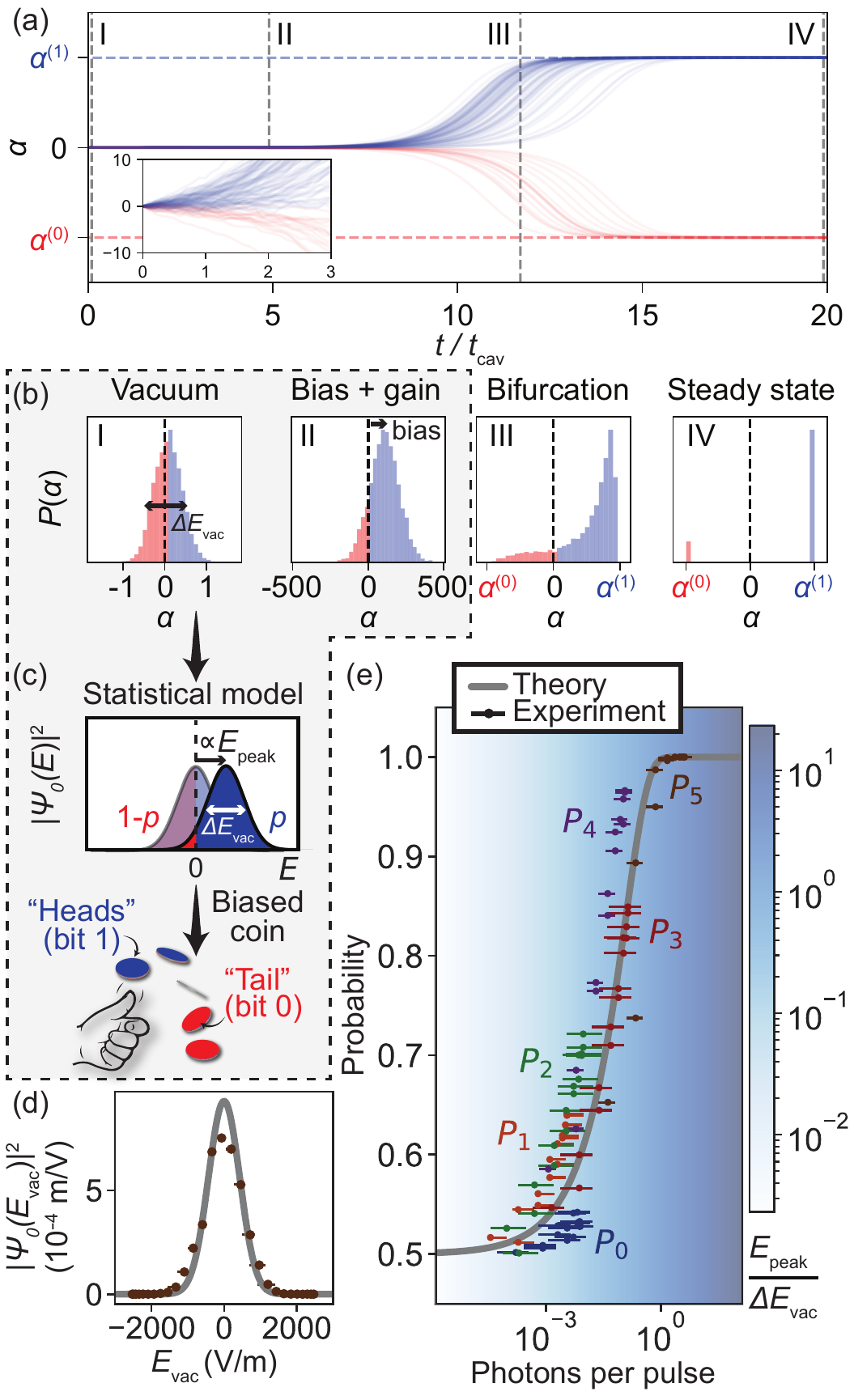}
    \caption{\textbf{Biased coin flips from the quantum vacuum.} (a) Stochastic trajectories of OPO amplitude. Time is normalized to the cavity lifetime $t_\text{cav}$. Inset: Zoom-in of the trajectories at early times. (b) Four-step model evolution of the probability distribution, highlighting the importance of early times in the determination of the bias-probability relationship. (c) Statistical model (from the bias and gain step in (b)) of the biased OPO as a biased coin flip, where the parameter $p$ is determined by the positive-value area under the ground state wavefunction density $|\Psi_0(E)|^2$ shifted by $E_\text{peak}$. (d) Reconstructed wavefunction density $|\Psi_0(E)|^2$ of the electromagnetic ground state. (e) Transition between purely random and deterministic regimes of the OPO. Transition occurs around half a photon per pulse, or equivalently, a peak bias field commensurate with the amplitude of vacuum fluctuations. The grey curve is the theoretical prediction without any fitting parameters, which comes from Equation~\ref{eq:prob-bias} with $\phi=0$, and $b_0 = 1/2$. For readability, only every other angle data point with probability $p>0.5$ is shown.}
    \label{fig:stat-model}
\end{figure}

\section*{Experimental demonstration of a photonic $p$-bit in a biased degenerate optical parametric oscillator}

We now experimentally demonstrate the system schematically depicted in Figure~\ref{fig:concept}(b,c): a weakly biased OPO, realizing a tunable $p$-bit in the photonic domain. A simplified schematic of our OPO experimental setup is shown in Figure~\ref{fig:expt}(a). The setup consists of three main parts: (1) an ultrafast degenerate OPO in a bow-tie cavity configuration (blue shaded area); (2) homodyne detection to measure the OPO phase (orange shaded area); (3) a delay and control line of the bias field to tailor the OPO probability distribution (green shaded area). A dual-wavelength laser system provides the pump ($2\omega$ = 780~nm, red line), the LO (local oscillator) and bias pulses ($\omega$ = 1560~nm, orange line). A quasi-phase-matched nonlinear crystal is placed between two spherical mirrors of the cavity. Through the interaction of the pump with the nonlinear crystal, a signal builds up with frequency $\omega$. The phase of the generated signal is detected by means of homodyne detection with the LO signal. This way, the pulse train generated by the OPO naturally provides a random bit stream. 

True randomness from vacuum fluctuations is realized as long as the signal in the cavity decays to the vacuum level between two consecutive samples, ensuring statistical independence of bits (this is achieved by modulating the pump at 10~kHz: see SM, Section~S4). With the bias field blocked (realizing an unbiased degenerate bistable Hamiltonian, similar to $H_0$ in Figure~\ref{fig:concept}(a)) we measured random bits with a uniform binomial distribution $\mathcal{B}_{0.5}$, and verified that the random bits pass a suite of statistical tests~\cite{rukhin2001statistical} (see SM, Section~S4).

We first demonstrate a continuously tunable \pbit~by controlling the bias field amplitude, realizing a binomial distribution $\mathcal{B}_p$. The intensity and time delay ($\Delta\tau_\text{bias}$) of the bias pulses are controlled by a set of attenuating filters, polarization optics, a spatial filter, and delay line. A combination of waveplates and polarizers enable the fine tuning of the bias amplitude. This provides a total attenuation of 90-125~dB (measured with respect to a bias input of $\sim114$~mW). In this first experiment, the bias signal is temporally synchronized with the pump ($\Delta\tau_\text{bias}=0$). 

The measured mean bit value estimated over $N=20,000$ samples (corresponding to probability $p$) is shown in Figure~\ref{fig:expt}(b), where the bias field amplitude sets the \pbit~probability. In our implementation, the bias field amplitude is set by a fixed attenuation (corresponding to the maximum power $P_i$) and fine-tuned by rotating the half waveplate angle, such that the bias power is given by $P_i\sin^2 \left( 2\theta\right)$. We fitted our experimental data to the functional form from Equation~\ref{eq:prob-bias}, showing an excellent agreement with our model. Specifically, this agreement indicates that the bit probabilities change in response to bias field attenuation as expected from theory.

By changing the attenuation level of the bias field, we were able to traverse the continuous space between perfect randomness and determinism. Segments of the bit stream statistics are shown in Figure~\ref{fig:expt}(c), showing a clear transition from a slightly skewed binomial distribution $\mathcal{B}_p$ with $p\approx0.5$ ($P_0$) to purely deterministic outcomes ($P_5$). We also verify that the bias does not introduce unwanted correlations in the bit stream with further statistical analysis described in Section~S4 of the SM. Importantly, the realized \pbit~exhibits the main properties necessary for its use in probabilistic computing platforms~\cite{camsari2019p, chowdhury2023full}: statistical independence of consecutive samples and continuous tunability of the probability distribution. Since the \pbit~outcome is in the optical domain, it could be further processed by an optical analog processor~\cite{shen2017deep, wetzstein2020inference}.

\section*{Observing vacuum fluctuations with photonic \pbits}

We now demonstrate that the probability control observed in Figure~\ref{fig:expt}(b) is the result of bias fields at the level of vacuum fluctuations. This allows us to elucidate the physics of the interplay between vacuum fluctuations and weak bias fields in the early time evolution of an OPO during symmetry breaking above threshold. To do so, we begin by explaining the key physical predictions revealed by our theoretical model. Figure~\ref{fig:stat-model}(a) shows sample trajectories simulated using stochastic differential equations derived from a rigorous quantum mechanical model of a biased OPO. The OPO steady state considered here contains $|\alpha_0|^2 = 10^{10}$ photons in a pulse, while the bias field contains just $b^2 = 0.64$ photons per pulse. Although the system starts in the ground state, the small bias field results in an uneven distribution of stochastic trajectories into the two steady states at $\pm\alpha_0$. 

Examining the probability distribution of fields at four characteristic times (marked I-IV) during evolution from the initial condition allows us to develop a four-step model of how the biased distribution develops:

\textbf{I. Initial vacuum fluctuations:} At $t=0$, there are no signal photons, but only vacuum fluctuations with a Gaussian distribution centered at zero field. In the absence of any bias, the two halves of the electromagnetic ground state are amplified away from the origin, resulting in an equal $(p= 0.5)$ distribution of phases.

\textbf{II. Bias and gain:} In the early stages of time evolution, the system is simultaneously impacted by parametric gain, a coherent bias field, and Gaussian white noise. During this period, the field probability distribution remains Gaussian. The bias shifts the mean away from zero, while the parametric gain amplifies both the mean and variance toward larger photon numbers. The inset of Figure~\ref{fig:stat-model}(a) shows a close up of this noisy gain and bias regime. Even though the field at this stage is small relative to its future steady state value, the final probability distribution is essentially determined by the field distribution at this stage. The statistical model presented in Figure~\ref{fig:stat-model}(c) relies on this observation. 

\textbf{III. Bifurcation:} As the signal field grows, saturation occurs as second harmonic generation of the signal field into the pump frequency acts as form of nonlinear absorption. During this stage, the Gaussian probability distribution created during the initial gain stage bifurcates into two lobes. It is clear from the time domain plots that at this stage, trajectories in a given phase will not cross over into the other phase.

\textbf{IV. Steady state:} Final evolution toward the steady state occurs as amplification and absorption become equal, and all fields converge onto one of the two possible steady state phases. At this stage, the two states exist with probabilities $p$ and $1-p$, with $p$ determined by stage 2. 

Our experimental data is quantitatively consistent with these theoretical predictions. The experiments shown in Figure~\ref{fig:expt}(b) are realized with $\approx 3\times10^{-5}-4$ photons per pulse (depending on the total attenuation level). In Figure~\ref{fig:stat-model}(e), the measured probability from various bias power levels $P_i$ (and waveplate angle $\theta$) is plotted against the estimated number of photons per pulse, matching our theory without any fitting parameter. Our theory not only accounts for the bias-probability relationship faithfully (as seen in Figure~\ref{fig:expt}(b)), but also accurately predicts the absolute level of bias fields required to influence the probability. Our theory and experiments both indicate that this critical bias field is on the order of one photon per pulse, as could be expected intuitively.

Equivalently, the critical bias peak field is comparable to the amplitude of the vacuum fluctuations $\Delta E_\text{vac}$ of the bias mode. As shown in the "bias + gain" phase of Figure~\ref{fig:stat-model}(b), early stages of the OPO evolution play a critical role in picking the steady state of weakly-biased OPOs. This observation allows us to extrapolate the following simple, but intuitive statistical model of the phase measurement process in a biased OPO: The measured phase is determined by the interference of the bias field $E_\text{peak}$ with a Gaussian random scalar field $E_\text{vac}$ with zero mean and standard deviation $\Delta E_\text{vac}$. The skew in the probability distribution requires a peak field on the order of $\Delta E_\text{vac}$. Therefore, the probability parameter $p$ is given by the area under the positive field side of the ground state wavefunction density $|\Psi_0(E_\text{vac})|^2$ displaced by $E_\text{peak}$, as shown in Figure~\ref{fig:stat-model}(c). 

This simple model, in agreement with our theory, also indicates the origin of the randomness in our experiment, and all other OPO-based random number generators: statistical fluctuations of the bias field in the early stages of field amplification. In this perspective, the OPO cavity serves as an amplifier of the original random seed generated in early time evolution of the weakly biased OPO. Interestingly, in low gain regimes characteristic of most experimental realizations, including ours, the probability distribution is approximately independent of the gain (as shown in Equation~\ref{eq:prob-bias} and in Section~S2 of the SM). This allows us to map the ground state wavefunction density $|\Psi_0(E_\text{vac})|^2$ to the bias-sensitivity of the probability $dp/db$ (shown in Figure~\ref{fig:stat-model}(d)). The reconstructed $|\Psi_0(E_\text{vac})|^2$ closely matches the expected distribution from quantum optics $|\Psi_0(E_\text{vac})|^2 \propto \exp \left( - E^2 / \left(2 \Delta E_\text{vac}^2\right) \right)$. This result is another strong indication that biased vacuum fluctuations are at the origin of the randomness in our experiment.

\begin{figure*}
    \centering
    \includegraphics[scale = 0.7]{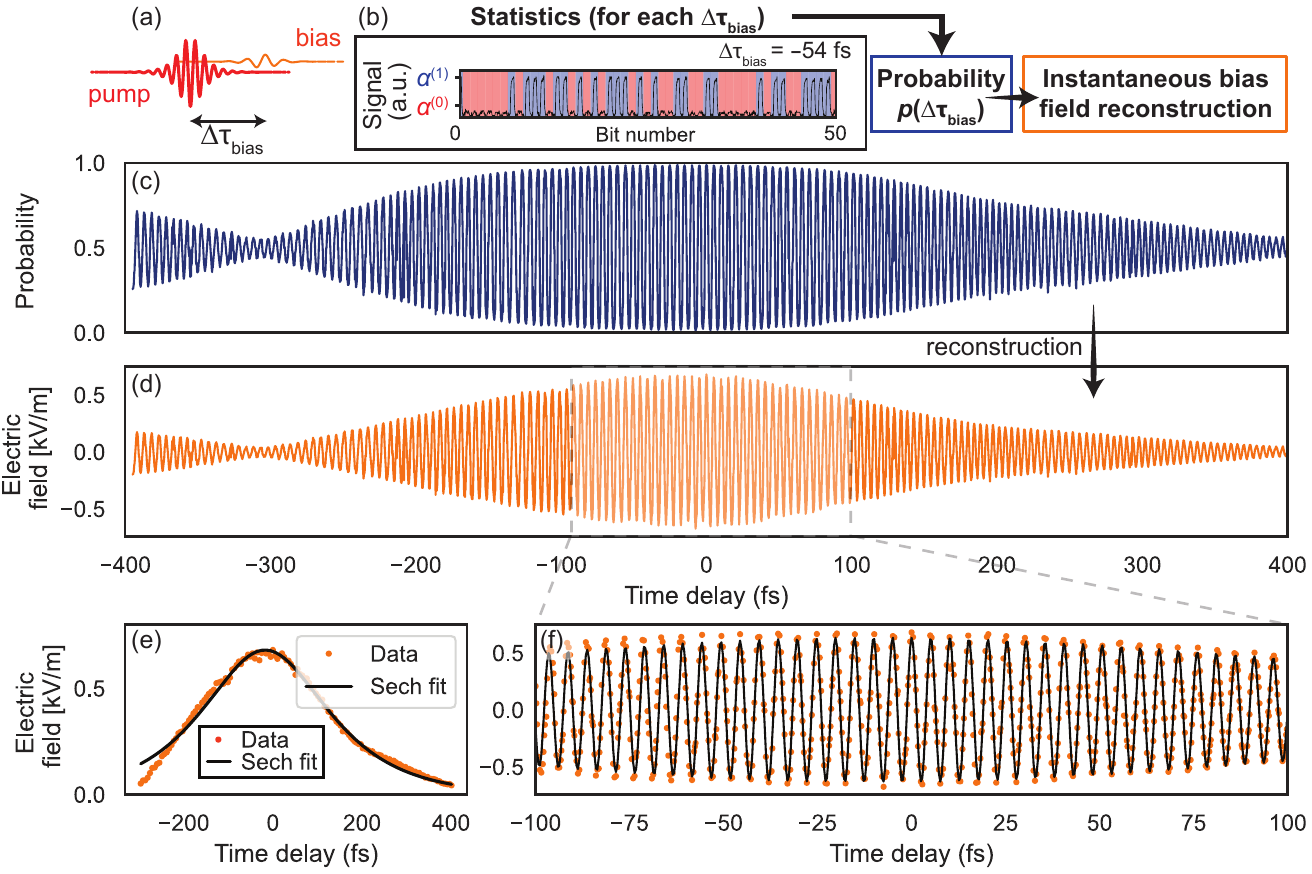}
    \caption{\textbf{Measuring sub-photon-level ultrashort pulses.} (a) Schematic of the bias control in this experiment, tuning the time delay $\Delta\tau_\text{bias}$ between the pump and bias. (b) Field reconstruction method: for each $\Delta\tau_\text{bias}$, we measure statistics of the biased OPO, extract the value of the probability $p$, and use it to reconstruct the instantaneous bias field. (c) Measurement of the probability as a function of the time delay $\Delta\tau_\text{bias}$. (d) Corresponding reconstructed bias field. (e) Extracted field envelope and corresponding $\text{sech}$ fit. (f) Central part of the pulse and corresponding $\text{sech}$ fit.}
    \label{fig:piezo-sweep}
\end{figure*}

\section*{Temporal reconstruction of sub-photon-level fields with photonic \pbits}

So far, we have demonstrated that the probability distribution of phases in a biased OPO is highly sensitive to sub-photon level bias fields in a manner which agrees quantitatively with theory. In this section, we show how the probability's sensitivity to the phase bias can be directly used to sense the temporal dependence of sub-photon fields (fields such that the mean number of photons in the pulse is less than one). 


A schematic of our sub-photon field sensing experiment is shown in Figures~\ref{fig:piezo-sweep}(a,b). The bias field is again strongly attenuated (down to an intensity corresponding to less than one photon per pulse) and time-delayed with respect to the pump. The time delay between the two pulses $\Delta\tau_\text{bias}$ is controlled and measured with a piezoelectric inertia actuator. At each $\Delta\tau_\text{bias}$, statistics of the OPO phase are recorded (over $N=20,000$ samples, an example of which is shown in Figure~\ref{fig:piezo-sweep}(b)). By inverting the bias-probability relationship from Equation~\ref{eq:prob-bias}, one can estimate the time-dependent bias field: $b\left(\Delta\tau_\text{bias}\right) \propto \text{erf}^{-1} \left( 2p\left(\Delta\tau_\text{bias} \right) - 1 \right)$, where $\text{erf}^{-1}$ is the inverse of the erf function. 

In our experiment, the reconstructed field $b\left(\Delta\tau_\text{bias}\right)$ (shown in Figure~\ref{fig:piezo-sweep}(d)) reproduces the main characteristics of the bias' temporal lineshape (mainly, its carrier frequency and pulse width). From the field lineshape, we can extract the pulse width and center wavelength (data and fits shown in Figure~\ref{fig:piezo-sweep}(e,f)). Our analysis yields estimated values for the main lobe of $\sim220\pm6$~fs and $\sim1520\pm0.2$~nm respectively, in good agreement with nominal values of the laser we used for this experiment (more details can be found in Section~S3 of the SM). We observe an asymmetric side lobe on the left side of Figure~\ref{fig:piezo-sweep}(d), which we attribute to the presence of a chirp in the bias or OPO signal (see Section~S3 of the SM).

In the generalized form of Equation~\ref{eq:prob-bias} we used for the field reconstruction, $b\left(\Delta\tau_\text{bias}\right)$ is actually the time-overlap between the bias and OPO signal at early times (in the first few cavity round trips, corresponding to the second step of the model shown in Figure~\ref{fig:stat-model}(b)). We present in Section~S2 of the SM further justification of this reconstruction method. In our experiment, the OPO signal at early times is much shorter (in time domain) than the bias, which explains why the reconstructed field approximates the bias field lineshape. Therefore, we also expect the reconstructed lineshape to depend on the OPO signal bandwidth and center frequency, as we further elucidate in complementary experiments shown in Section~S3 of the SM. 

With this proof of concept, one could envision using this reconstruction method as a form of statistical and phase-sensitive weak signal sensing technique. In this method, statistics of the system's response yield information on the bias field and early-stage OPO signal, and could therefore be used to reconstruct either in more general settings, with prior knowledge of the other and an additional deconvolution step. 



\section*{Discussion and conclusion}

We have presented a method in which we inject a vacuum-level bias in nonlinear driven-dissipative systems to sense their vacuum fluctuations via statistical measurement. In quantum systems, such as the biased OPO we have utilized as a proof of concept, our results shed light on the interplay between vacuum-level bias fields and vacuum fluctuations of the electromagnetic fields. Our results also elucidate the importance of early-stage dynamics in nonlinear systems in setting steady-state probability distributions. Taken together, the methods proposed in this work suggest the use of vacuum-level bias fields as a probe of transient dynamics in nonlinear driven-dissipative systems by observing their influence on steady-state dynamics.



Specifically, the dynamic range of our weak signal sensing method spans over five orders of magnitude and its sensitivity is determined by the number of measured statistical samples $N$. The maximum sensitivity of our measurements (shown in Figure~\ref{fig:expt}-\ref{fig:piezo-sweep}), corresponding to the linear part of the erf function, is on the order of $\sqrt{\frac{\pi}{4N}}$, corresponding to roughly $6\times10^{-3}$ photons per pulse for the $N$ used in these experiments. Sensitivity could be improved by collecting more statistical samples.

Our results also demonstrate the use of quantum fluctuations as a resource for generating random numbers with a tunable probability distribution. The results shown in Figure~\ref{fig:expt} constitute an experimental demonstration of a photonic \pbit~where randomness is realized and controlled by the interference of the bias and vacuum fields. An anticipated extension of this work will be to incorporate this photonic \pbit~into a computing platform to realize photonic probabilistic computing~\cite{camsari2019p}. This can be realized by leveraging the extensive body of work on photonic Ising machines~\cite{wang2013coherent, marandi2014network,mcmahon2016fully,inagaki2016coherent,yamamoto2017coherent,pierangeli2019large,roques2020heuristic,prabhu2020accelerating,vadlamani2020physics,mohseni2022ising}. Specifically, in coherent Ising machines, OPOs are coupled in space or time using time multiplexing and/or measurement-and-feedback schemes emulating many-body Hamiltonians~\cite{marandi2014network,mcmahon2016fully,inagaki2016coherent, honjo2021100}.

The realized \pbit~is, to the best of our knowledge, the first demonstration of a \pbit~in the photonic domain. Currently, our \pbit~can generate 10,000 bits of arbitrary binomial distributions $\mathcal{B}_{p(b)}$ per second, where the probability parameter $p(b)$ is controlled by the amplitude, phase, or time delay of the weak bias field $b$. The main limitations come from the control electronics and cavity mode lifetime~\cite{marandi2012all}. To increase the repetition rate of our photonic \pbit, one could resort to bistable systems in integrated photonics~\cite{okawachi2016quantum} or complex mode interference in laser cavities~\cite{kim2021massively}. By using fast modulation of the bias field delay, phase, or amplitude, photonic \pbits~with $>$~GHz repetition rates could be realized in the near future. The photonic \pbit~bias-sensitivity ($dp/db$) can be adjusted by controlling the amplitude of the bias field or, equivalently, the effective volume of the vacuum mode (e.g., by changing the pulse width in time, or its waist in the nonlinear crystal). 

The proposed framework also offers a general approach to realizing \pbits~in nonlinear driven-dissipative quantum systems, harnessing zero-point fluctuations as a source of noise. Realizing our approach in multistable systems (e.g., in integrated arrays of nanolasers~\cite{parto2020realizing, altug2005photonic}) paves the way to \pbits~with extreme bandwidth (by leveraging spatial multiplexing), emulation of complex many-body Hamiltonians (by engineering couplings between \pbits), and non-trivial \pbit~topologies (by using Hamiltonians $H_0$ with higher-order symmetries). We envision that such systems will exhibit complex dynamics with potential applications in combinatorial optimization and lattice quantum chromodynamics simulations.

\section{Authors contributions}
C.~R.-C., Y.~S., and M.~S. conceived the original idea. C.~R.-C. and Y.~S. built the experimental setup with contributions from S.~C., G.~V., E.~K., and S.~K.~ C.~R.-C., Y.~S., J.~S., and S.~C. acquired and analyzed the data. J.~S. developed the theoretical and numerical tools with contributions from C.~R.-C. and N.~R.~ J.~D.~J. and M.~S. supervised the project. The manuscript was written by C.~R.-C., Y.~S., and J.~S. with inputs from all authors. 

\section{Competing interests}
The authors declare no potential competing financial interests.

\section{Data and code availability statement}
The data and codes that support the plots within this paper and other findings of this study are available from the corresponding authors upon reasonable request. Correspondence and requests for materials should be addressed to C.~R.-C. (chrc@mit.edu) and Y.~S. (salamin@mit.edu).

\section{Acknowledgements}
The authors would like to thank Adam Heiner, Joseph Mastron, and Alois Wiesboeck from Toptica for their help in setting up the dual-wavelength femtosecond laser. The authors also acknowledge administrative support from Joshua Freedman. The authors acknowledge stimulating conversations with Shiekh Zia Uddin, Christina Spägele, Rumen Dangovski, Di Luo, Samuel Kim, Zin Lin, and Erich Ippen. Y.~S. acknowledges support from the Swiss National Science Foundation (SNSF) through the Early Postdoc Mobility Fellowship No.~P2EZP2-188091. Research was sponsored by the Army Research Office and was accomplished under Cooperative Agreement Number W911NF-18-2-0048.


\bibliography{bibliography.bib}

\end{document}